\documentclass[iop,revtex4]{emulateapj}
\usepackage{float,epsfig}
\usepackage{lscape}
\usepackage{natbib}
\usepackage{color}
\usepackage{amsmath}

\newcommand{\bats}{Swift J0042.6+4112}

\def\cxo{{\it Chandra\/}}

\def\hst{{\it {\it HST}\/}}
\def\swift{{\it {\it Swift}\/}}

\def\spitzer{{\it Spitzer\/}}

\def\xmm{{\it XMM-Newton\/}}
\def\nustar{{\it NuSTAR\/}}

\def\msun{$M_\odot$}
\def\h2{H$_2$}

\def\nh{$N_{\rm H}$}
\def\ergcms{erg~cm$^{-2}$~s$^{-1}$}
\def\ergl{erg~s$^{-1}$}
\def\arcsec{\mbox{$^{\prime\prime}$}}
\def\arcmin{\mbox{$^\prime$}}

\def\aap{A\&A}
\def\apj{ApJ}
\def\apjl{ApJL}
\def\apjs{ApJS}
\def\aj{AJ}
\def\mnras{MNRAS}
\def\araa{ARA\&A}
\def\pasj{PASJ}
\def\nat{Nature}
\def\spose#1{\hbox to 0pt{#1\hss}} 
\def\simlt{\mathrel{\spose{\lower 3pt\hbox{$\sim$}}
        \raise 2.0pt\hbox{$<$}}}
\def\simgt{\mathrel{\spose{\lower 3pt\hbox{$\sim$}}
        \raise 2.0pt\hbox{$>$}}}

\begin{document}

\shorttitle{\nustar\ \& \swift\ Observation of \bats}

\title{Identification of the Hard X-ray Source Dominating  the $E >$ 25 keV Emission of the nearby galaxy M31}

\author{M.~Yukita\altaffilmark{1,2}, A. Ptak\altaffilmark{2,1}, A.~E.~Hornschemeier\altaffilmark{2,1},
D.~Wik\altaffilmark{1,2}, T.~J.~Maccarone\altaffilmark{3}, K.~Pottschmidt\altaffilmark{4,5}, A.~Zezas\altaffilmark{6,7}, V.~Antoniou\altaffilmark{6}, R.  Ballhausen\altaffilmark{8}, B.~D.~Lehmer\altaffilmark{9}, A.~Lien\altaffilmark{4,5}, B. Williams\altaffilmark{10}, 
F.~Baganoff\altaffilmark{11}, P.~T.~Boyd\altaffilmark{2},T.~Enoto\altaffilmark{12,13}, J.~Kennea\altaffilmark{14}, K.~L.~Page\altaffilmark{15}, and Y.~Choi\altaffilmark{16}}

\altaffiltext{1}{The Johns Hopkins University, Homewood Campus, Baltimore, MD 21218, USA}
\altaffiltext{2}{NASA Goddard Space Flight Center, Code 662, Greenbelt, MD 20771, USA}
\altaffiltext{3}{Department of Physics \& Astronomy, Texas Tech University, Lubbock, TX 79409, USA}
\altaffiltext{4}{CRESST, Department of Physics, and Center for Space Science and Technology, UMBC, Baltimore, MD 21250, USA} 
\altaffiltext{5}{NASA Goddard Space Flight Center, Code 661, Greenbelt, MD 20771, USA} 
\altaffiltext{6}{Harvard-Smithsonian Center for Astrophysics, 60 Garden Street, Cambridge, MA 02138, USA}
\altaffiltext{7}{Physics Department, University of Crete, Heraklion, Greece}
\altaffiltext{8}{Dr. Karl-Remeis-Sternwarte and Erlangen Centre for Astroparticle
Physics, Sternwartstr. 7, 96049 Bamberg, Germany}
\altaffiltext{9}{Department of Physics, University of Arkansas, 226 Physics Building, 835 West Dickson Street, Fayetteville, AR 72701, USA}
\altaffiltext{10}{Department of Astronomy, University of Washington, Seattle, WA 98195, USA}
\altaffiltext{11}{MIT Kavli Institute for Astrophysics and Space Research, 77 Massachusetts Ave, Cambridge, MA 02139, USA}
\altaffiltext{12}{The Hakubi Center for Advanced Research, Kyoto University, Kyoto 606-8302, Japan}
\altaffiltext{13}{Department of Astronomy, Kyoto University, Kitashirakawa- Oiwake-cho, Sakyo-ku, Kyoto 606-8502, Japan}
\altaffiltext{14}{Department of Astronomy \& Astrophysics, The Pennsylvania State University, 525 Davey Lab, University Park, PA 16802, USA}
\altaffiltext{15}{Department of Physics \& Astronomy, University of Leicester, Leicester  LE1 7RH, UK}
\altaffiltext{16}{Steward Observatory, University of Arizona, 933 North Cherry Avenue, Tucson, AZ 85721, USA.}

\begin{abstract}
We report  the identification of  a bright hard X-ray source dominating the
M31 bulge above 25 keV from a simultaneous  \nustar-\swift\ 
observation.  We find that this source is the counterpart to Swift
J0042.6+4112, which was previously detected in the \swift\
BAT All-sky Hard X-ray Survey.  This \swift\ BAT source had been suggested to be the combined emission 
from a number of point sources;  our new observations have identified a single X-ray source from 0.5 to 50 keV as
the counterpart for the first time. 
In the 0.5--10 keV band, the source  had been classified as an
X-ray binary candidate in various \cxo\ and \xmm\ studies; however,
since it was not clearly associated with Swift
J0042.6+4112, the previous E $<$ 10 keV
observations did not generate much attention. This source has a spectrum with  
a soft X-ray excess ($kT\sim$ 0.2 keV) plus a hard spectrum
with a power law of $\Gamma\sim$ 1  and  a cutoff around
15--20 keV, typical of  the spectral characteristics of accreting
pulsars. Unfortunately,  any potential  pulsation 
  was undetected in the \nustar\ data, possibly due to insufficient photon statistics. The existing deep \hst\ images
exclude high-mass ($>$3 \msun) donors at the location of this source.
The best interpretation for the nature of this source is
an X-ray pulsar with an intermediate-mass ($<$3 \msun) companion or a symbiotic X-ray binary. We  discuss other possibilities
in more detail.

\end{abstract}

\keywords{stars: neutron --- (stars:) pulsars: general  --- galaxies: individual (M31) --- galaxies: bulges  --- X-rays: binaries}

\section{Introduction}
Thanks to the sensitivity and spatial resolution of \nustar,
we can investigate the E$>$10 keV properties of nearby galaxies in detail.
We now know that starburst and normal galaxies
have X-ray spectra which drop quickly above 10 keV \citep[]{wik14,lehmer15,yukita16}.
Since a soft spectrum above 10 keV is observed in ultraluminous X-ray sources (ULXs), black hole (BH) binaries in their intermediate accretion state, 
and Z-sources (a subclass of neutron star low-mass X-ray binaries), we
conclude that some combination of these types of sources
is likely dominating the integrated galaxy
spectra at harder energies ($>$10 keV).  Individual
resolved sources in starburst and normal galaxies similarly show high-energy cutoffs
around 5--15 keV \citep[e.g.,][]{church12,bachetti13,lehmer13,walton13,walton14}. 

We expect starburst galaxies, whose specific star formation rates (sSFRs) are high, to be dominated by short-lived high-mass X-ray binaries, and  that more quiescent galaxies with lower sSFRs have a contribution from low-mass X-ray binary (LMXB) systems.
To isolate the E $>$ 10 keV spectral properties  of the LMXB population
that are related to older stellar populations,
we have observed the M31 bulge, which shares  basic properties (kinematics, stellar populations, etc.) with
early-type galaxies.

\begin{figure*}
\begin{center}
\includegraphics[angle=0,width=0.9\textwidth]{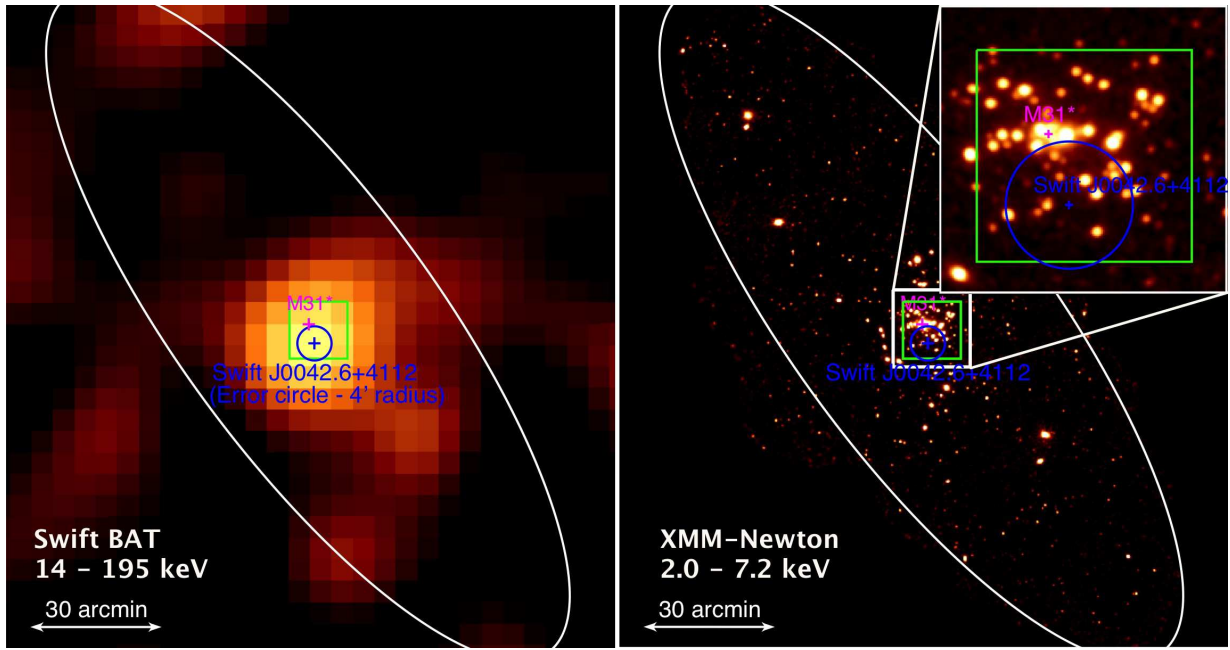}
\end{center}
\caption{Left: The 70 month {\it Swift} BAT 14--195~keV image of M31.
The white ellipse shows the optical extent ($D_{25}$)  of M31.
The magenta cross depicts the location of M31*. The blue cross indicates the location of \bats\ with a 4\arcmin\ 
radius position error circle. The green box indicates the FoV of our {\it NuSTAR} observation.
Right: \xmm\ 2.0--7.2~keV image of 
M31 (from http://heasarc.gsfc.nasa.gov/docs/xmm/gallery/esas-gallery/xmm\_gal\_science\_m31.html)
    The inset image is the magnified view of the the bulge region showing 
a number of resolved point sources within the BAT error circle in the 2.0--7.2 keV band with \xmm.}
\label{f:xmmbat}
\end{figure*}

M31 was previously  detected  at  hard energies
 in the \swift\ BAT (14--195 keV) all-sky survey \citep{baumgarner13} with 7$\sigma$ significance.  
 The BAT flux is dominated by a single source, \bats, located 6\arcmin\ away from 
 M31's dynamical center with $F_{\rm X}$ = 9 $\times$ 10$^{-12}$ \ergcms\ 
 in the 14--195 keV band.  Its classification is listed as unknown
in the BAT catalog. 
 \citet{revnivtsev14} investigated the broadband  3--100 keV
 spectrum of the integrated M31 galaxy
 emission based on \textsl{RXTE}, \swift\ BAT, and {\it INTEGRAL} data and suggested that \bats\ represents
 the total emission from the M31 galaxy above 20 keV,  but  $<$6\% of the total X-ray (3 -- 100 keV) luminosity for M31.  
The 2--10 keV  luminosity  of the source, extrapolated from its hard X-ray luminosity, would be $>$ 5$ \times$ 10$^{38}$ \ergl. Although this luminosity is bright enough to be detected with \xmm\ or \cxo, there is no unique counterpart found 
 in \xmm\ or \cxo\ images \citep{revnivtsev14}. Hence,  \citet{revnivtsev14} surmised it is a collection of very faint sources rather than a single bright point source.  

Recently, we have obtained  simultaneous \nustar-\swift\ observations
of the M31 bulge, which detected and resolved roughly  20 X-ray point
sources above 10~keV in the bulge region.  
We note that all the other resolved X-ray sources with \nustar\
 will be reported in a follow-up paper (D. Wik et al. 2017 in-prep).

In this paper, we report the discovery of  a {\it single}  \nustar\
point-source counterpart to \bats, which completely dominates the bulge region 
 at E $>$ 25 keV.  
We  describe the \nustar\ and \swift\ data and data reduction in Section 2.    In Section 3, the X-ray data analysis of \bats\ is performed.  Section 4 investigates its possible optical counterparts.
We discuss the nature of this source in Section 5. We adopt a
distance to M31 of 784 $\pm$ 13 kpc \citep{stanek98}, for which 1\arcsec\
corresponds to 3.8 pc. Unless noted otherwise, quoted uncertainties  correspond to 90\% confidence intervals for one interesting parameter.

\section{Observations \& Data Reduction}
The M31 bulge was observed with \nustar\ on 2015 October 12--14 ({\tt obsid} 50101001002, PI: Yukita) for 98.5 ks.
The data were then processed from level 1 to level 2 using the {\tt nupipeline} script available in HEASoft version 6.17 with
CALDB version 20151008.  This observation suffered from strong background flares during passes through the South Atlantic Anomaly, and we manually
removed contaminated periods from the good time intervals (GTIs), resulting 
in a final exposure of 92.2 ks.  
Source spectra were extracted using a 45\arcsec\ radius circular aperture,
and  response files (ARFs and RMFs) were
created using the {\tt nuproducts} script.  The background spectra were extracted using a source-free region
near the source in the same observation.   The \nustar\ FMPA and FMPB
spectra and corresponding response files were co-added
using the {\tt addascaspec} script, and then the final source spectrum was grouped to achieve at least one count in each bin  \citep[see][for discussion about grouping \nustar\ data]{wik14b}.

\swift\ observed the M31 bulge on 2015 October 13--14  ({\tt obsids} 00081682001 and 00081682002), simultaneous  with
 part of the \nustar\ observation, 
for a total XRT exposure of 17 ks.  The XRT spectra
were extracted using a 45\arcsec\ radius circular region, and the ARFs were created using the {\tt xrtmkarf} tool.
The relevant RMFs for the observations were obtained from CALDB.  
We note that within our \swift\ observations there was exactly one XRT source consistent with the \nustar\
counterpart to the \swift\ BAT source;
there are no other XRT sources within the \nustar\ aperture for the counterpart. 
Similarly to the \nustar\ data, the background spectra were  extracted using the source-free region near the source 
in the same observations.
The \swift\ XRT spectra and responses were also combined using the {\tt addascaspec} script. 
The co-added XRT spectrum was grouped to achieve at least one count in
each bin. Spectral analysis was performed in {\it XSPEC} version
12.9.0 using the C-statistic.

\begin{figure*}
\begin{center}
\includegraphics[angle=0,width=0.95\textwidth]{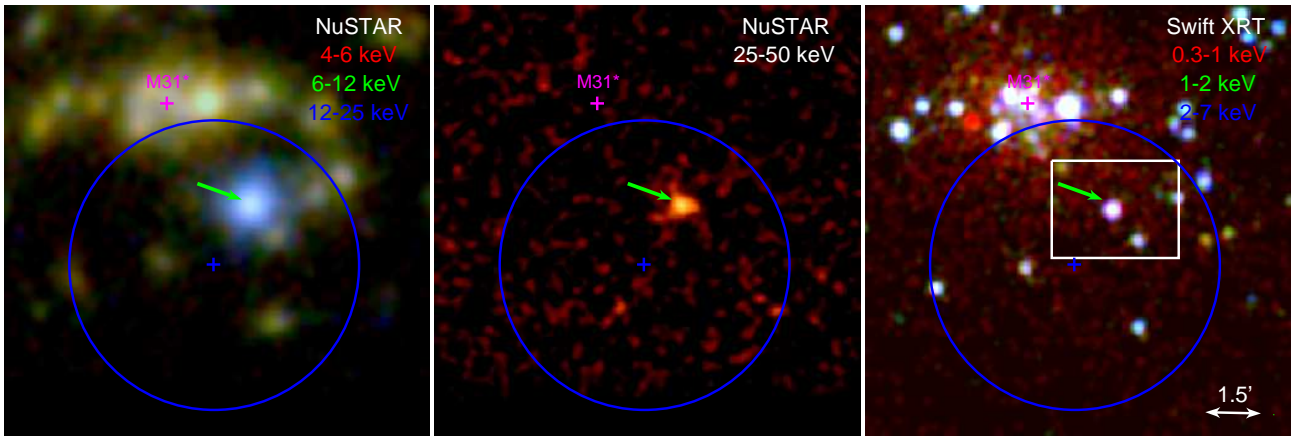}
\end{center}
\caption{Left:  \nustar\ color image of the M31 bulge region.  The blue cross depicts the \swift\ BAT position of \bats. 
There are several \nustar\ point sources identified within the 4\arcmin\ radius position error circle (blue circle).
There is one source  that is harder (bluer) compared to other point sources (indicated by the green arrow). The image was smoothed
with a Gaussian kernel with  $\sigma$=10\arcsec. Middle:  \nustar\ 25--50 keV image of the bulge region.
The blue source in the left panel is the only source that appears in this band.  We identify this blue source as the
 counterpart of \bats.  The image was smoothed
with a Gaussian kernel with  $\sigma$=7\arcsec. Right:  \swift\ color image (co-added the two observations, the total of 17 ks exposure) of the bulge region. There is a bright source in the \swift\ image 
at the location of the \nustar\ counterpart of \bats. The image was smoothed
with a Gaussian kernel with  $\sigma$=5\arcsec.  The white box depicts the region shown in 
Figure~\ref{f:cxoimage}.}
\label{f:bulgeimage}
\end{figure*}

\section{\nustar\ counterpart of \bats}
The \swift\ BAT all-sky survey  \citep{baumgarner13} identified a hard X-ray source along the line of sight of M31 (see the left panel of Figure~\ref{f:xmmbat}),
and this is  the only source listed in the 70-month catalog within the $D_{25}$ of M31.
At $E<$10 keV, using higher angular resolution observations such as \xmm\ and \cxo, 
there are quite a number of X-ray point sources detected (see  the right panel of Figure~\ref{f:xmmbat}) in the M31 bulge \citep[i.e.,][]{kong02,stiele11,hofmann13,barnard14}.
However, no single 0.5-10 keV X-ray point source has been constrained as a counterpart of  \bats.

\subsection{Source Identification}
Figure~\ref{f:bulgeimage} shows the \nustar\ (left and middle panels) and \swift\ (right panel) images of the M31 bulge region. 
Running the {\it CIAO} wavdetect tool on the \swift\ 0.3--7 keV image, we detected 10 sources within the BAT 91\% position error 
circle \citep{baumgarner13} indicted as a blue (4\arcmin\ radius) circle in Figure~\ref{f:bulgeimage}.  We note that the detection limit of
our \swift\ observations is $\sim$10$^{36}$~\ergl\ in the 0.3 -- 7 keV band;
however, the simultaneity of the \swift\ observations is of great benefit to anchor the archival (and
therefore nonsimultaneous) \cxo\ and \xmm\ observations to guard against the effect of variability.  
These 10 \swift\ sources in our observation were all listed in the
\cxo\ catalog \citep{barnard14}, which
contains more than 50 point sources within the error radius,
reaching down to a luminosity limit of $\sim$ 5 $\times$ 10$^{34}$~\ergl\ in the 0.3--10 keV band. 

The left panel of Figure~\ref{f:bulgeimage} shows that several \swift\ point sources seen in the right panel of  Figure~\ref{f:bulgeimage}  
are also apparent in the
\nustar\ 4--25 keV color image.  
We point out that there is a distinctive hard source appearing in blue within the position error circle of \bats\ in the \nustar\ image. 
This blue source is also seen in the \nustar\ 25--50 keV image (middle panel).  
In fact, this is the only source apparent in the FoV in this harder 25 -- 50 keV band. 
Therefore, this source is likely to be the sole \nustar\ counterpart of \bats.
This  source is also clearly detected in the better spatial resolution  \swift\ 0.3--7.0 keV image. 

\begin{figure*}
\begin{center}
\includegraphics[angle=0,width=0.95\textwidth]{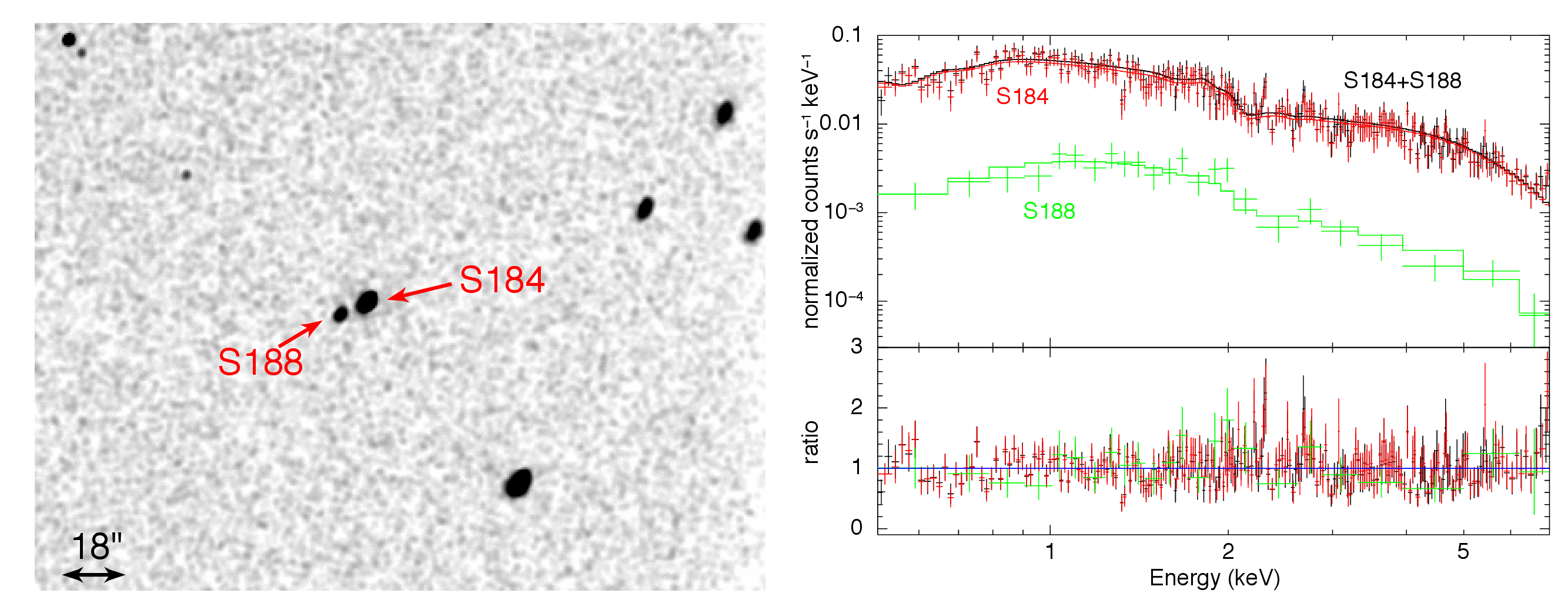}
\end{center}
\caption{Left: The \cxo\ image taken in 2012 (obsid 13826) for the location of the \nustar\ counterpart of \bats. In the higher angular resolution \cxo\  image, the point source detected with \nustar\ (PSF $\sim$ 58\arcsec\ HPD) and \swift\ 
(PSF $\sim$ 18\arcsec\ HPD) is clearly
resolved into two point sources, S184 and S188 \citep{barnard14}. Right:  The \cxo\ spectra, best-fit models, and fit ratios of S188 (green), 
S184 (red) and their sum (black). S184 is an order of magnitude brighter than S188, dominating
the total spectrum of the two sources completely.  The \cxo\ spectral properties in 2012 of these two sources are
consistent with those fit to the \swift\ and \nustar\ data from Sept 2015. The spectra are rebinned to achieve at least 3$\sigma$ or are grouped in sets of 30 bins for
display purpose.}
\label{f:cxoimage}
\end{figure*}

To obtain a better source position, we also examined the archived \cxo\ data and published source catalogs \citep{hofmann13,barnard14}. 
The high-resolution ($\sim$0.5\arcsec) \cxo\  data (left panel of Figure~\ref{f:cxoimage}) reveals that
there are two X-ray point sources \citep[S184, and S188;][]{barnard14} separated by $\sim$8\arcsec\ at the location of the \nustar\ counterpart of \bats, although
these sources are unresolved and appear as a single source in both the \nustar\ and \swift\ data. 

To determine which of the \cxo\ sources is the  counterpart of \bats,
we performed spectral analysis  using the 2012 \cxo\
observation  (obsid 13826, 37 ks exposure; one of the longest exposures) and compared
it to the \swift\ observation taken in 2015. 
We applied an absorbed power-law model for S184 and obtained a power-law index of $\sim$1.4
with C-stat of 522 (with 410 dof).  This model is statistically disfavored with null probability of $<$10$^{-4}$
(no simulations were as bad as the model fit). Then, we applied an absorbed power law plus disk blackbody,
 which was chosen as a canonical X-ray binary spectral model, and we
  defer detailed discussion of the spectral properties to the next subsection.
We note that  S188 did not contain enough counts ($\sim$240 counts in the 0.5 -- 7.0 keV band) to fit complicated models.
Therefore, we fitted absorbed 
power-law or disk-blackbody models only.
Table~\ref{t:cxospec} lists the fitting results.  
In 2012, the observed flux of  S188  in the 0.5--7.0 keV band   was less than 10\% of S184.  
When summing the spectra of both sources, the fitted parameters listed in Table~\ref{t:cxospec}
are consistent with those of S184, having a $T_{\mathrm in} \sim$ 0.2 keV and power-law index of $\Gamma$ $\sim$ 1.
The fitting results confirm that S184 dominated the spectra in the 0.5--7.0 keV as also shown in
the right panel of Figure~\ref{f:cxoimage}.
We point out these best-fit parameter values are consistent with the \swift\ spectrum (for the sum of S184 and S188) taken
in 2015, namely, having a $T_{\mathrm in} \sim$ 0.2 keV and power-law index of $\Gamma$ $\sim$ 0.8 (see Table~\ref{t:cxospec}), 
as well as the \nustar\ spectra exhibiting a flat power-law slope ($\Gamma$ =1.2$\pm$0.2) in the 3.0 -- 7.0 keV band. 
 S184 dominates the soft X-ray spectral properties in 2012, and its spectral properties are consistent with 
both S184 and S188 sources combined in the 2015 \swift\ data. 
In addition, the light curves for the two \cxo\ sources  published by \citet{hofmann13}
show that S188 was always fainter than S184 during 2006 -- 2012. 
Hence, we assume that  S184 also dominates during our \swift\ observations and
determine that S184 is likely the \cxo\ counterpart for \bats.
With the better \nustar\ PSF compared to \swift\ BAT's angular resolution,  \bats\  is identified
as a single X-ray point source,  previously known as S184 or CXOM31 J004232.0+411314 \citep{kong02}
 with \cxo\ coordinates of   $\alpha_{2000}$= 00:42:32.072 and $\delta_{2000}$ = $+$41:13:14.33 \citep{barnard14}, 
from 0.5 to 50 keV for the first time. We note that the statistical uncertainty of \nustar\ position (1 $\sigma$) of this source is $\sim$0.25\arcsec. 
There is no robust measurements of systematic uncertainty of the \nustar\ position, and we conservatively use 2.5\arcsec\ (1 pixel scale). Therefore,
we estimate a \nustar\ position uncertainty of $\sim$2.5\arcsec\ for \bats.

 Interestingly, S184 is one of the brighter point sources ($>$10$^{37}$ \ergl) in the M31 bulge and has been detected with
various X-ray telescopes (e.g., {\it Einstein}: \citealt{van97,trinchieri91}; {\it ROSAT}: \citealt{primini93,supper01}; \cxo: \citealt{kong02,kaaret02,hofmann13,barnard14}; 
\xmm: \citealt{pietsch05,stiele11}).
Its flux and spectral variabilities have been reported \citep[e.g.][]{kong02}. The source has been interpreted as an
X-ray binary candidate in M31 without further discussion of the nature of this source.   We now investigate the nature of \bats\ 
 to examine what type of point sources dominate M31 at harder energies.

\begin{deluxetable*}{cccccccccc} 
\tabletypesize{\scriptsize}
\tablecolumns{10} 
\tablewidth{0pc} 
\tablecaption{0.5--7.0 keV spectral analysis results\label{t:cxospec}} 
\tablehead{     
  \colhead{}   & \colhead{\nh}    & \colhead{$T_{\mathrm in}$}   & \colhead{Norm}     
 &  \colhead{}    & \colhead{Norm}   & \colhead{$f^{obs}$(0.5--7.0 keV)}  &    \colhead{} & \colhead{}  & \colhead{ Null}\\
 \colhead{SRC}   & \colhead{(10$^{20}$~cm$^{-2}$)}    & \colhead{(keV)}   & \colhead{DBB}     
 &  \colhead{$\Gamma$}    & \colhead{PL}   & \colhead{(\ergcms)}  &    \colhead{C-stat} & \colhead{dof} & \colhead{ prob}
  }
\startdata 
\multicolumn{10}{c}{\cxo\ 2012 Jun 6 (obsid 13826) } \\
\hline
S184  &  7 & \nodata  & \nodata & 1.41$\pm$0.05& (1.4$\pm$0.6)$\times$10$^{-5}$ &  9.2$\times$10$^{-13}$ &   522 & 410 &  $<$10$^{-4}$\\ 
S184   & 7 &   0.20$\pm$0.03 &13$^{+13}_{-6}$& 0.92$\pm$0.11 &  (8.5$\pm$1.1)$\times$10$^{-5}$ &  1.0$\times$10$^{-12}$ &  383 & 408 & 0.6503\\ 
S188   & 7 &  1.35$^{+0.32}_{-0.22}$ & 0.0010$^{+0.0008}_{-0.0005}$ & \nodata &  \nodata& 5.9$\times$10$^{-14}$& 100 &  149 &  0.5934\\ 
S188   & 7 &   \nodata & \nodata &1.52$\pm$0.19 &  (1.2$\pm$0.2)$\times$10$^{-5}$  & 6.5$\times$10$^{-14}$& 98 &  149 &  0.7615\\ 
S188+S184 & 7 & 0.20$\pm$0.03 & 14$^{+15}_{-7}$ & 0.99$\pm$0.10&  (9.8$\pm$1.6)$\times$10$^{-5}$& 1.1$\times$10$^{-12}$  &  391 & 412&  0.6083 \\ 
\hline
\multicolumn{10}{c}{\swift\ 2015 Sep 13--14 (obsids 00081682001 and 00081682002) } \\
\hline
S188+S184 & 7 &  0.21$^{+0.08}_{-0.06}$&  12$^{+50}_{-9}$& 0.80$^{+0.33}_{-0.41}$ &  7.4$^{+3.6}_{-3.2}$$\times$10$^{-5}$& 1.1$\times$10$^{-12}$ & 224 &  263&
$>$0.9999 \\ 
\enddata
\tablenotetext{}{Note -- Due to small number counts in the spectra, the \nh\ value is fixed to the Galactic column density of 7$\times$
10$^{20}$~cm$^{-2}$ \citep{dickey90}. PL: power law.  DBB: disk blackbody.  Both \cxo\ and \swift\ spectra were grouped to achieve at least 1 count per bin, and C-stat is used for fitting. Null probability is calculated from the {\it XSPEC} {\rm GOODNESS} command using  the Anderson-Darling statistic test.
A null probability around 0.5 indicates the observed spectrum is produced by the model.}
\end{deluxetable*}

\subsection{\nustar\ \& \swift\ Spectral Analysis} 
In this section, we explore the 0.5--50 keV spectral properties of \bats\ 
with \nustar\ and \swift. 
We limit the joint spectral analysis to the simultaneously taken  \nustar\ and \swift\  observations, as we see flux variabilities
of this source (see Tables \ref{t:cxospec} and \ref{t:2010spec}).
Since  both \nustar\ and \swift\ spectra include emission contributed
from S188,  we take its contribution into account by including the fit model of S188 from Table~\ref{t:cxospec}.  Two S188 models
are indistinguishable, but we use an absorbed
 disk-blackbody model as it is a slightly better description of the observed spectrum (null probability of 0.6).  We fixed the
parameter values, except for allowing the normalization
to vary by up to 30\%.  We note, however, that the fit results are
consistent within the errors when we do not include the S188 component.
Therefore, contamination by emission from S188 has no appreciable impact on the conclusions we make here.

\begin{figure}
\begin{center}
\includegraphics[width=0.48\textwidth]{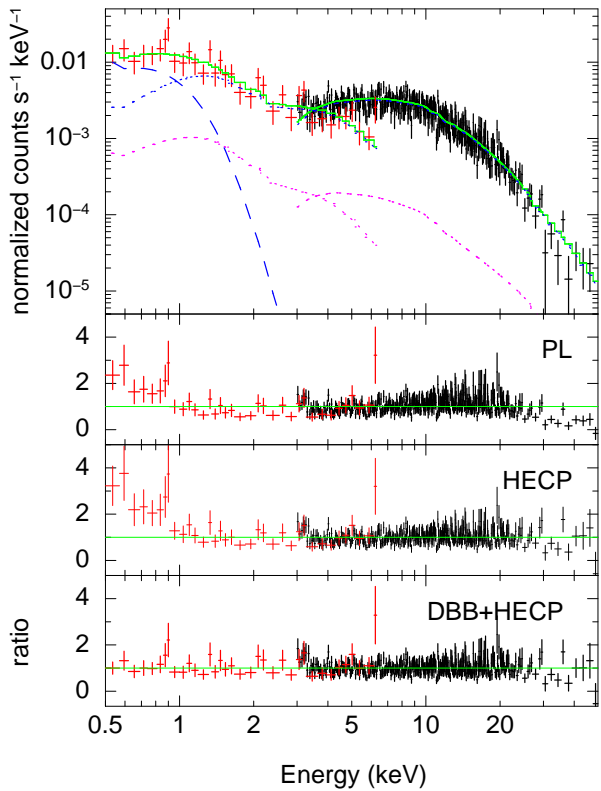}
\end{center}
\caption{Top: The \nustar\ (black) and \swift\ (red) spectra of  \bats/S188  combined are shown.
The best-fit model (disk blackbody + high energy cutoff power law and a power law) for the combined spectra are shown in green.
The model components for S184 and S188 are shown in blue and magenta, respectively. 
Dash lines depict the disk-blackbody model. The power law components are
shown with dot lines. Second top to bottom: The fit ratio to the data for power law, high energy cutoff power law, and disk blackbody
plus high energy cutoff power law, respectively.  These illustrate there is an excess in the softer band and a cutoff at harder energies.
The spectra were binned in the plots here for
display purposes only.}
\label{f:jointspec}
\end{figure}

Initially, we fit S184 (the \bats\ counterpart) with a simple absorbed power-law model. 
 The photon index and null probability from this model were $\sim$1.2 and 0.0002, respectively (see
Table~\ref{t:xspec}).  This model is disfavored statistically.   
Inspecting the residuals, this power-law model deviates from the data
below 1 keV and above 30 keV (see Figure~\ref{f:jointspec}), suggesting that the spectrum has curvature.   
Hence, we proceeded with a high-energy cutoff power-law model, 
 often used for accreting X-ray pulsars \citep[i.e., strong magnetic field neutron
star binary systems;][]{mullar13}. 
This model was marginally accepted with the null probability of $\sim$0.12.
We also apply a broken power-law model  to compare with the BAT power-law slope.
Both models fit reasonably well (null probability of $\sim$0.21), including the data above 20 keV,
demonstrating that the spectrum steepens at higher energies.
The photon index above the break energy for the
broken power-law model is $\Gamma_2$ =  2.47$^{+0.46}_{-0.45}$, which agrees within the errors with the power law index of
$\Gamma$ = $2.97^{+0.72}_{-0.53}$ in the 14--195~keV band
obtained from the BAT 70-month survey \citep{baumgarner13}.

To better fit the spectrum below 1 keV, we added a disk-blackbody component.
This additional soft component models the spectrum below 1 keV well
with $T_{in}$ = 0.2 keV, consistent with parameter values listed in
Table~\ref{t:cxospec}  and reduces C-stat by  50 for one
  additional parameter. The null probability is also improved from $\sim$0.12 to $\sim$0.75. 
We note that the absorption was fixed to the Galactic value, as there were not enough counts in the \swift\ spectrum
to constrain both \nh\ and the disk-blackbody component.  
We also applied the disk-blackbody plus broken power-law model; which fits equally well (see Table~\ref{t:xspec}).
 We note that the
additional soft component does not impact  the broken power law or 
high energy cutoff power law model parameters significantly. We  also note
that the disk-blackbody component flux is less than 15\% of the total flux in the 0.5--7.0 keV band.
Table~\ref{t:xspec} tabulates the fitting results.

The 0.5--50 keV spectral analysis results suggest that \bats\ possesses similar X-ray 
spectral properties to accreting X-ray pulsars that have a soft X-ray excess \citep[e.g.,][]{hickox04}. 
The observed flux in 0.5--50 keV for the best-fit model (disk
blackbody plus high-energy cutoff power law) is
5.5 $\times$10$^{-12}$ \ergcms\ (the unabsorbed flux does not differ, due
to the relatively low \nh\ value).  
Assuming that this object is located within M31, the corresponding luminosity is  4.0 $\times$ 10$^{38}$~\ergl\ in the 0.5--50 keV band.
The brightest known accreting X-ray pulsars typically achieve similar luminosities (e.g., LMC X-4; \citealp{hung10}, SMC X-1; 
\citealp{neilsen04}, RX J0059.2-7138, \citealp{hughes94}; M82 X-2, NGC7793 P13  and  NGC 5907 ULX1 for exceptionally bright examples; \citealp{bachetti15,furst16,israel17,israel16}).

The \swift\ BAT 70-month averaged flux in the 14--195 keV band is 9.65$^{+2.95}_{-2.61}$ $\times$ 10$^{-12}$ \ergcms\ \citep{baumgarner13}.
We estimate the 14--195~keV flux for our observation by extrapolating the disk-blackbody plus a high energy cutoff power-law model and
obtain  3.9 (3.8 for broken power law) $\times$ 10$^{-12}$ \ergcms, which is about 40 \% of the BAT 70-month flux.  
We also fit the disk-blackbody plus a high-energy cutoff power-law model
to the \nustar, \swift\ XRT, and \swift\ BAT 70-month averaged
spectrum, with  the \swift\ BAT spectrum scaled to take variability into account.  
This fit is consistent with the results presented in Table~\ref{t:xspec}, with the BAT 70-month
spectrum having a normalization 2.2 times higher for the 
high-energy cutoff power law component.
This suggests that the hard X-ray flux was a factor of 2 fainter in 2015 compared to the average flux 
during 2006--2012, or that there were additional sources that varied within the BAT
source region. 

The BAT 70-month light curve does not show strong variability \citep{baumgarner13}, but the signal-to-noise ratio is not high, and the statistical uncertainty may be large. 
 We note that the 0.5--7.0 keV flux did not change in observations between 2012 and 2015 (Table~\ref{t:cxospec}); however, the 
\cxo\ light curve between 2006 and 2012 shows flux variability more than factor of 2  in the 0.2--10 keV band 
\citep[see Figure B1, source 75 of ][]{hofmann13}.   We also confirmed that the 0.5--7.0 keV flux in 2010 was about 2--4 times dimmer 
(see Table~\ref{t:2010spec}).

Concerning the possibility that other sources are contributing to the BAT
source flux, there are faint sources located outside
of the \nustar\ 45\arcsec\ radius aperture, which were undetected
in the \nustar\ data at harder energies.
Some of these faint sources could contribute to the flux listed in the BAT catalog.
We examined how much hard X-ray emission (in the 25--50 keV band) comes from \bats\  compared to
the total \nustar\ FoV in our \nustar\ observation. We note that the background for the \nustar\ FoV
was estimated using the {\tt nuskybgd} tool \citep[see][for details]{wik14b}. We found that no more than 15\% of emission
is contributed from sources other than \bats\ in the \nustar\ 25--50 keV band, and it is unlikely that undetected faint point sources largely contribute to
the BAT flux.

\begin{deluxetable*}{cccccccccc} 
\tabletypesize{\scriptsize}
\tablecolumns{10} 
\tablewidth{0pc} 
\tablecaption{0.5--50~keV \nustar-\swift\ joint spectral analysis results\label{t:xspec}} 
\tablehead{     
  \colhead{Model$^a$}   & \colhead{\nh}    & \colhead{$T_{\mathrm {in}}$}   & \colhead{}     
 &  \colhead{$E_{\mathrm {cutoff}}/E_{\mathrm {br}}$$^b$}    & \colhead{$E_{\mathrm {efold}}/\Gamma_{2}$$^b$}   & \colhead{$f^{obs}$(0.5--50 keV)}  &    \colhead{} & \colhead{}  & \colhead{Null}\\
 \colhead{+S188}   & \colhead{(10$^{20}$ cm$^{-2}$)}    & \colhead{(keV)}   & \colhead{$\Gamma/\Gamma_{1}$$^b$}     
 &  \colhead{(keV)/(keV)}    & \colhead{(keV)/--}   & \colhead{(\ergcms)}  &    \colhead{C-stat} & \colhead{dof}& \colhead{prob$^c$}
  }
\startdata 
PL & 7.0$^{+1.3}_{-0.0}$ & \nodata & 1.15$\pm$0.03  & \nodata & \nodata & 7.0$\times$10$^{-12}$  & 1207 & 1143 & 0.0002\\ 
BKNPL & 7.0$^{+0.9}_{-0.0}$ &\nodata& 0.99$^{+0.06}_{-0.05}$ &18$^{+2}_{-3}$ & 2.47$^{+0.46}_{-0.45}$ &  5.4$\times$10$^{-12}$&   1100 & 1141 & 0.2076 \\ 
HECP &  7.0$^{+0.8}_{-0.0}$& \nodata & 1.00$^{+0.04}_{-0.05}$& 17$^{+2}_{-3}$& 19$^{+5}_{-4}$ & 5.4$\times$10$^{-12}$&   1100 &1141 &  0.1211\\ 
DBB+HECP   & 7.0 & 0.19$^{+0.06}_{-0.04}$ &0.86$^{+0.08}_{-0.06}$& 14$^{+3}_{-2}$ & 19$\pm4$ &  5.5$\times$10$^{-12}$ &  1050& 1140& 0.7498\\ 
DBB+BKNPL   & 7.0 & 0.19$^{+0.04}_{-0.06}$ &0.91$^{+0.05}_{-0.07}$ &   17$\pm2$&  2.31$\pm$0.31&   5.6$\times$10$^{-12}$& 1051 &   1140 & 0.8839\\ 
\enddata
\tablenotetext{a}{PL: power law, BKNPL: broken power law, HECP: high energy cutoff power law, DBB: disk blackbody}
\tablenotetext{b}{Parameters for a broken power law model}
\tablenotetext{c}{Null probability is calculated from the {\it XSPEC} {\rm GOODNESS} command using  the Anderson-Darling statistic test.
A null probability around 0.5 indicates the observed spectrum is produced by the model.}
\end{deluxetable*}

\subsection{Timing Analysis}
We performed timing analysis  on the \nustar\ data to in order to
search for spin and orbital periods. 
Detecting a pulsation would be strong evidence that the X-ray source is
a neutron star system.  Also, a relation between pulse period and orbital 
period gives some information such as mass transfer mechanisms \citep[i.e., Corbet diagram;][]{corbet86}.

 We binned the \nustar\  barycenter corrected lightcurve of \bats\
(combined FMPA and FMPB) using the 3.0--50 keV band.  We first fit a constant to the lightcurve binning by 7000 s, and obtained 
$\chi^2$/dof = 39.2/27, suggesting possible moderate flux variability on timescales of several
hours.  We note that the net count rate of both telescopes for a 45\arcsec\ 
aperture is $\sim$ 0.085 count s$^{-1}$ (the background is included). 
Using the barycenter corrected 3-50 keV events we then looked for any
periodicity between 0.1 and 10000s applying the epoch folding technique \citep{leahy83}.  We used 32 phase bins, and 
 roughly 18.6 million test periods were investigated for this period range. 
 However, we did not find significant signals besides the 5.8 ks  period of the satellite orbit.
Based on Equation 15 of \citet{leahy83} we determine an upper limit
for a possible pulsation amplitude of 0.08 counts s$^{-1}$ on a 99\%
confidence level.

We note that there will be another \nustar\ observation of this source simultaneously taken with \xmm\ 
during \nustar\ Cycle 2, and detailed long-term variability  including
the use of the archived \cxo, \xmm\, and \swift\ data will be
presented in a future paper. 

\begin{figure}
\begin{center}
\includegraphics[origin=c, angle=0,width=3.3in]{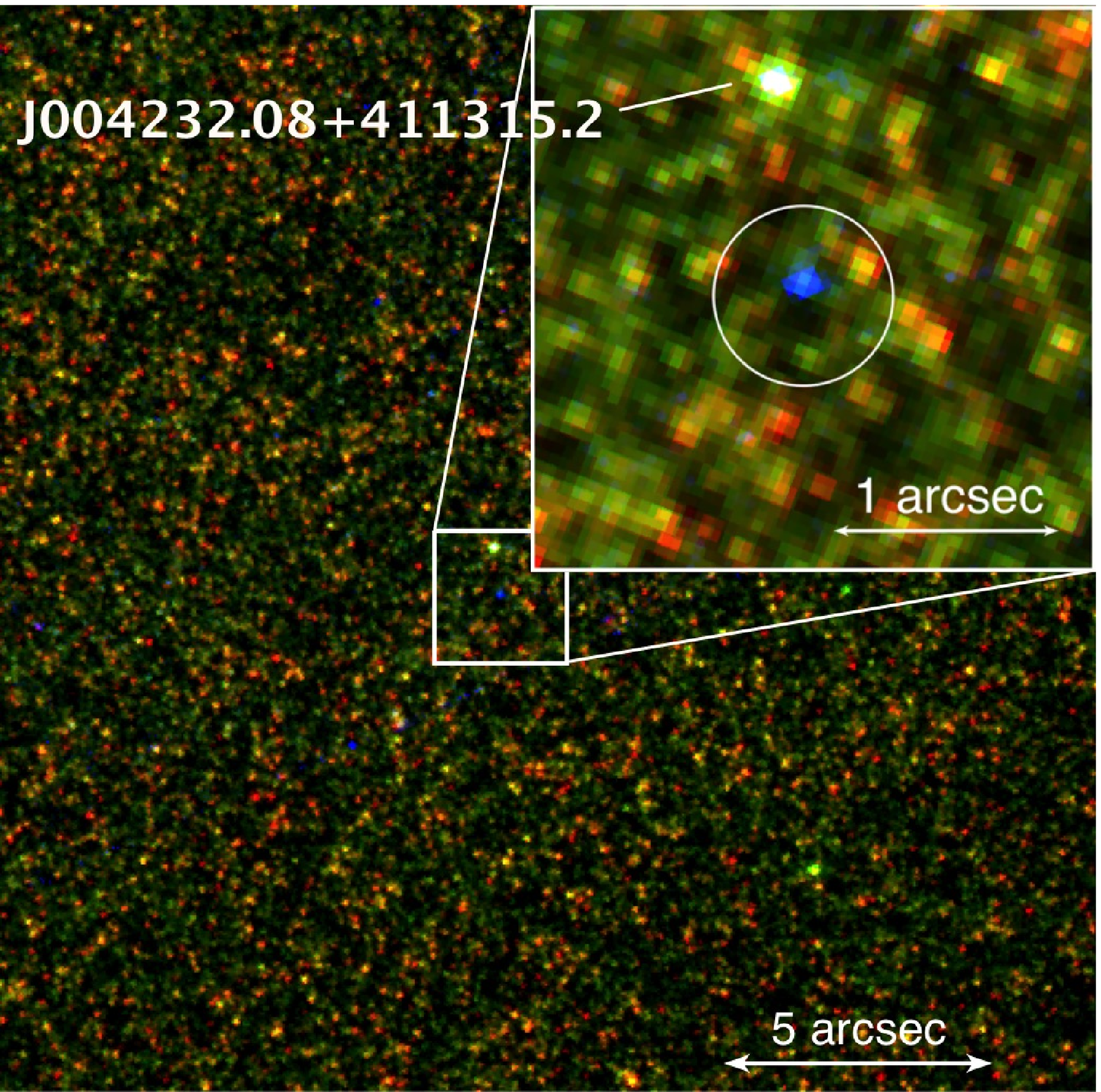}
\end{center}
\caption{The \hst\ color image (F336W [blue], F475W [green], and F814W [red]) of the region around S184, which likely is the \cxo\
counterpart of \bats. Numerous individual sources are resolved and detected with \hst. The inset shows the magnified image
of S184. The white circle (0.4\arcsec\ radius) depicts the \cxo\ position of \bats. There are 17 \hst\ 
sources within this circle detected by \citet{williams14}. The bright source J004232.08+411315.2 is used to register the \hst\ and
\cxo\ coordinates.}
\label{f:hstsrcs}
\end{figure}

\begin{figure*}
\begin{center}
\includegraphics[origin=c, angle=-90,width=3.4in]{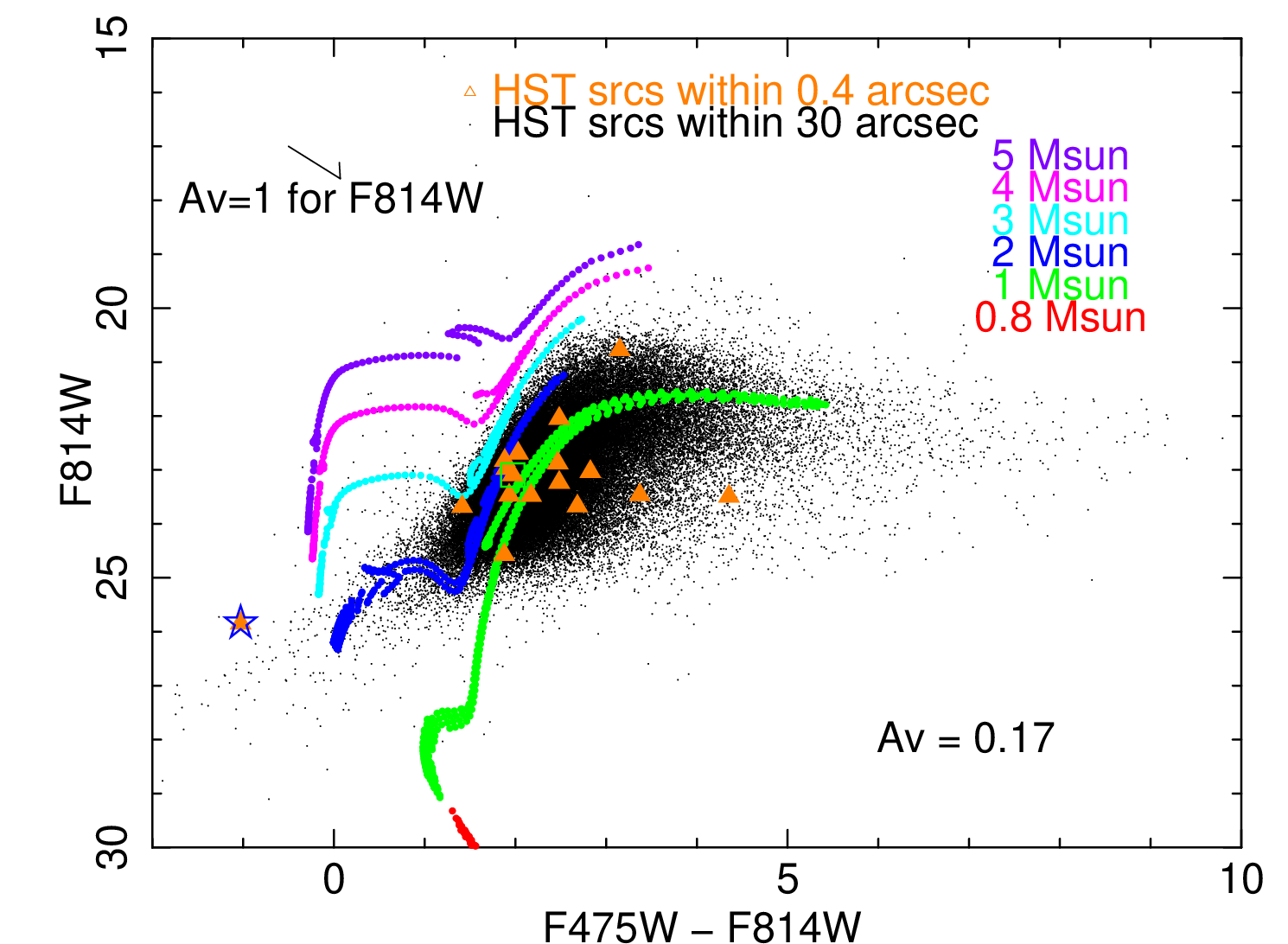}
\includegraphics[origin=c, angle=-90,width=3.3in]{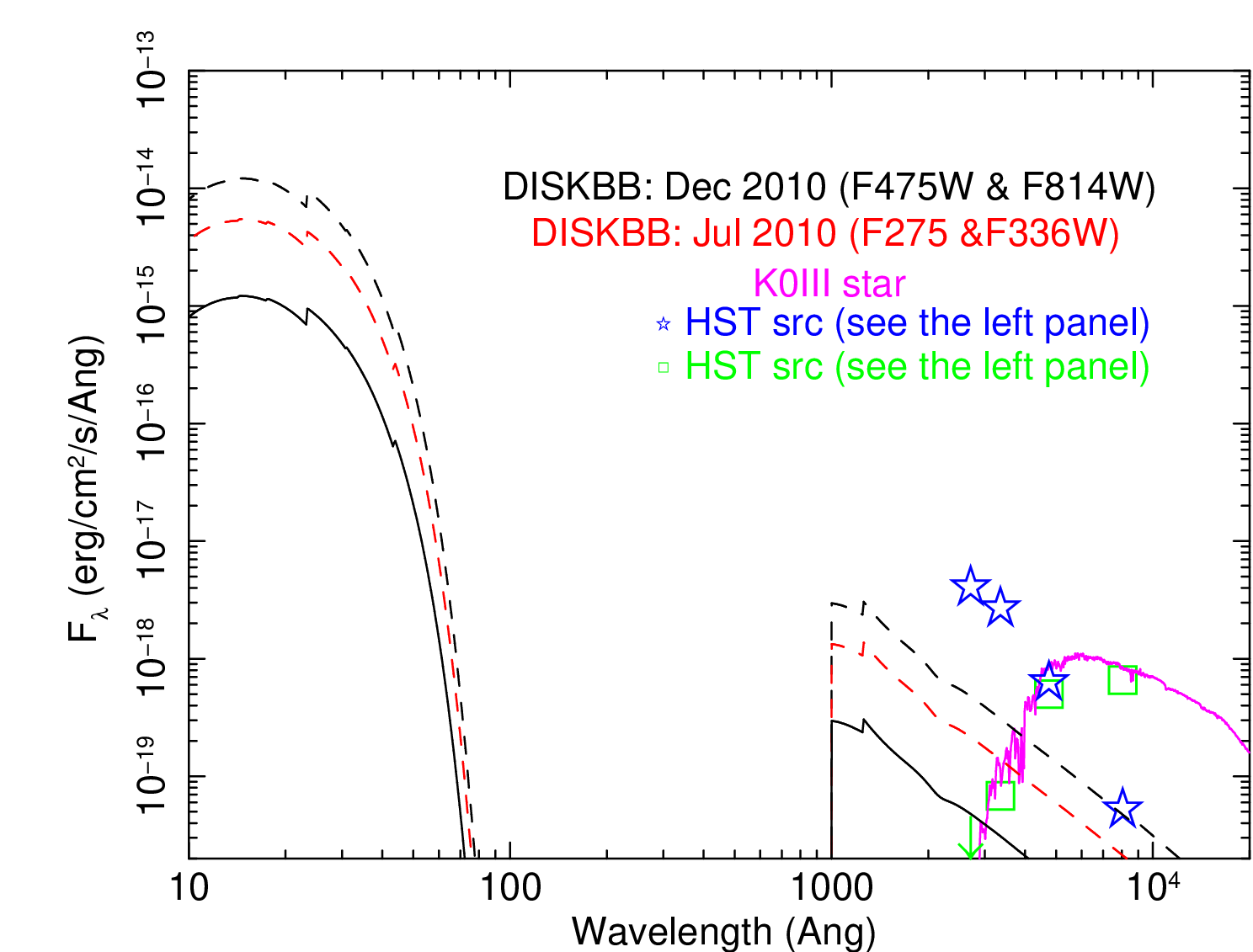}
\end{center}
\caption{Left: The color-magnitude diagram using the 
F475 and F814 filters for the \hst\ sources located within the 0.4\arcsec\ radius circle (orange triangles) and 30\arcsec\ radius (black dots)
 of the \cxo\ counterpart position of \bats. The stellar evolution tracks for 0.8, 1, 2, 3, 4 and 5~\msun\ stars are overplotted.
 No main-sequence stars are found as the \hst\ matches of \bats. Most detected sources are likely to be evolved intermediate-mass stars. 
 Two \hst\ matches whose SEDs are shown in the right are indicated with the open blue star and green square symbols.  Right: The
 disk-blackbody model for S184  obtained from 2010 July (red) and December (black) observation (see  Table~\ref{t:2010spec}).  The upper boundary (from normalization) is shown in dashed lines.  The blue stars and green squares  depict \hst\ photometry for two \hst\ matches as examples. It is unlikely
 that the \hst\ blue counterpart is the accretion disk.  A K0III spectrum from the Castelli and Kurucz Atlas \citep{castelli04} is shown in magenta as an example of evolved 
 intermediate-mass stars.   The spectra in the UV and optical regime are convolved with the extinction model by \citet{cardelli89} using $A_{\rm V}$ = 0.17. We note that the cutoff of the UV and optical spectra at 1000 Angstrom is artificial.}
\label{f:hst_counters}
\end{figure*}

\section{UV \& Optical counterparts}
The X-ray spectral properties  suggest that \bats\ is likely an accreting pulsar. 
In this section, we examine the UV to optical properties of the source to help determine its
 nature. 

The bulge of M31 is partially observed with \hst\  as part of the Panchromatic Hubble Andromeda Treasury \citep[PHAT,][]{dalcanton12} Survey. 
The area of the sky around \bats\ is covered with 4 filters (F275W, F336W, F475W and F814W).
The reduced images and photometry of individual sources are published by \citet{williams14},
which we use in our analysis below.

Figure~\ref{f:hstsrcs} shows an \hst\ three color (F336W, F475W, and F814W for blue, green,
and red, respectively) image at the location of  the \cxo\ source S184.
We adopted the source position published by \citet{barnard14}, which was registered to the LGS M31 Field 5 $B$-band image by \citet{massey06}.  
We checked for  any astrometry offset between the \hst\ and \cxo\ positions, using
the brightest optical source marked in Figure~\ref{f:hstsrcs}, J004232.08+411315.2 in \citet{massey06}.
The  position of J004232.08+411315.2 reported by \citet{massey06} agrees with the \hst\ position reported by \citet{williams14}  within  $\sim$0.06\arcsec. 
Therefore, further corrections to the astrometry were not necessary.

We searched for UV/optical counterparts of \bats\ within a  $\sim 3\sigma$
 \cxo\ error circle with  a radius of 0.4\arcsec\  \citep{barnard14} .
There are 17 \hst\ sources listed within this 0.4\arcsec\ radius in \citet{williams14}.
Utilizing the photometry  published in that work,
we constructed a color-magnitude diagram using the F814W and F475W
filter magnitudes as shown in the left panel of
Figure~\ref{f:hst_counters}.  
In this figure, we overplotted the PARSEC stellar evolutionary  tracks \citep{bressan12} for 0.8, 1, 2, 3, 4, and 5 \msun\ stars. 
We assumed solar  metallicity \citep{saglia10}, and 
the extinction was fixed at the Galactic value of $A_{\rm V}$ = 0.170 mag \citep{schlafly12}.
The observed column density in the M31 bulge area is relatively low \citep[$\sim$ log 21 cm$^{-2}$;][]
{braun09}.  However, high spatial extinction maps indicate there are the patchy dust lanes in the bulge
region \citep[e.g.,][]{li09,doug16}.    We investigated  the \spitzer\  8\micron\  map by
\citet{li09} and also created the F475W-F814W color map to look for potential dust lanes. 
We determined that the region around \bats\ is not likely
to have a high extinction and using a Galactic excitation value is adequate. 

Figure~\ref{f:hst_counters} suggests that all \hst\ sources within the 0.4\arcsec\ radius 
region except one are very similar to the surrounding stellar populations.
We also find that there are no high-mass ($>$3~\msun) stars within the 0.4\arcsec\ radius region.  
In addition,  no \hst\ matches are consistent with main sequence stars;
most of them are likely  evolved intermediate- and/or low-mass stars.
We note that there is a unique \hst\ blue source in the 0.4\arcsec\ region (the open blue star in Figure \ref{f:hst_counters}).
Fitting the 4-band SED using the Bayesian Extinction and Stellar Tool \citep{gordon16} suggests that it is not consistent
 with any model of stable evolutionary phases, suggesting that it is a 
 post-AGB and/or the x-ray source is contaminating the UV and optical flux.
We examined the F275W-F336W 
color and magnitude of this blue source and found that  it is likely a  hot post-horizontal branch star
at the distance of M31 as studied in \citet{rosenfield12}.  
The surface density for  hot post-horizontal branch stars at the location of \bats\  is lower than 0.01 arcsec$^{-2}$ \citep{rosenfield12};
therefore, the probability that such a star is found within the 0.4\arcsec\ radius circle is $<$0.005.   This is a
small value and makes it an interesting \hst\ match.

The optical emission associated with Galactic LMXBs are related to their accretion disks rather than their companions. 
We also checked whether or not  any of these optical matches  could be an accretion disk by comparing the
\hst\ photometric data to the best-fit disk-blackbody component.   
We first analyzed the \cxo\ and \xmm\ data (obsids 11840 and 0650560201), which are taken on similar epochs as the \hst\ PHAT data
(2010 July 24 for F275W and F336W, and 2010 December 25--26 for F475W and F814W) to measure the disk-blackbody
flux during those periods.
Similar to the previous section, 
we applied an absorbed power-law plus disk-blackbody model in the 0.5--7.0 keV band.
The results are given in Table~\ref{t:2010spec}.
 Due to the small number of counts in the spectra (especially the \cxo\ spectrum), we fixed the absorption to the Galactic value.
The \cxo\ observation in 2010 July did not constrain the disk component; therefore, we fixed the disk temperature at $T_{in}$ = 0.2 keV.
For the \xmm\ data taken in 2010 December, the spectrum contains both S184 and S188 as they were not resolved
in the \xmm\ observation. However,
we point out that the disk component parameters agree within the uncertainties  with and without
the S188 contribution using the listed values in Table~\ref{t:cxospec}. 
These spectral fitting results suggest that \bats\ was 2--5 times dimmer than in the Oct 2015 \nustar-\swift\
observation.  The disk component is barely detected.

The right panel of Figure~\ref{f:hst_counters}  shows two SED (F814W, F475W, F336W and F275W photometry data points) 
 examples of the \hst\ matches  with the best-fit
absorbed disk-blackbody models obtained from Table~\ref{t:2010spec} to see if the
UV/optical counterpart is related to the accretion disk. We note that the power-law component is
unlikely to be emitted in the UV-optical band; therefore, it is omitted.
The figure suggests that these two \hst\  matches are not likely due to
accretion disk flux.  For example, the open green square match has an observed optical emission that is higher than predicted
from the  disk model.  Moreover, its UV/optical emission is rather consistent with a K0III star spectrum, as expected from the
color-magnitude diagram.
Another example, the blue match with open blue star symbol, is not consistent with the accretion disk model either. 
We repeated the same exercise for the remaining sources and found that
no \hst\ sources are consistent with the accretion disk model. 
This suggests that the optical matches are not related to the accretion disk and are likely
stellar objects.

We note that it is possible that the system has a companion that is below the detection limit
of the \hst\ data (i.e., $<$ 2 \msun\ main-sequence star or a bit lower if it is an evolved star).

\begin{deluxetable*}{cccccccccc} 
\tabletypesize{\scriptsize}
\tablecolumns{10} 
\tablewidth{0pc} 
\tablecaption{0.5--7.0 keV spectral analysis results during 2010\label{t:2010spec}} 
\tablehead{     
  \colhead{}   & \colhead{\nh}    & \colhead{$T_{\mathrm {in}}$}   & \colhead{Norm}     
 &  \colhead{}    & \colhead{Norm}   & \colhead{$f^{obs}$(0.5--7.0 keV)}  &    \colhead{} & \colhead{} & \colhead{Null} \\
 \colhead{SRC}   & \colhead{(10$^{20}$~cm$^{-2}$)}    & \colhead{(keV)}   & \colhead{DBB}     
 &  \colhead{$\Gamma$}    & \colhead{PL}   & \colhead{(\ergcms)}  &    \colhead{C-stat} & \colhead{dof} & \colhead{prob}
  }
\startdata 
\multicolumn{10}{c}{\cxo\ 2010 July 20 (obsid 11840)} \\
 S184   & 7 &   0.20 &  2$^{+7}_{-2}$ &  1.26$^{+0.29}_{-0.35}$ & 8.3$^{+2.1}_{-2.8}$$\times$10$^{-5}$ & 6.3$\times$10$^{-13}$&   93 & 119 &  0.9261\\ 
\hline
\multicolumn{10}{c}{\xmm\ 2010 December 26 (obsid 0650560201) PN only} \\
S188+S184 & 7 &  0.20$^{+0.21}_{-0.17}$&  2$^{+20}_{-2}$& 1.22$^{+0.26}_{-0.46}$ &  2.7$^{+0.9}_{-1.3}$$\times$10$^{-5}$& 2.3$\times$10$^{-13}$ & 311 &  407 & 0.8961\\ 
\enddata
\tablenotetext{}{Note -- Due to small number counts in the spectra, the \nh\ value is fixed to the Galactic column density. PL: power law.  DBB: disk blackbody. 
Both \cxo\ and \xmm\ spectra were grouped to achieve at least 1 count per bin, and C-stat is used for fitting.  Null probability is calculated from the {\it XSPEC} {\rm GOODNESS} command using  the Anderson-Darling statistic test.
A null probability around 0.5 indicates the observed spectrum is produced by the model.}
\end{deluxetable*}

\section{Discussion}
We have identified a single hard X-ray ($>$25 keV) source within the error circle of \bats\ with \nustar.
In a simultaneous \swift\ observation, we also detected an X-ray point source at the location of this \nustar\ source
in the 0.5--7.0 keV band.  We have investigated the high
spatial resolution \cxo\ data, as well as the literature, and  pinpointed the location of this source to within $\sim$0.4\arcsec. 
In this section, we discuss the possible nature of \bats. 

First, we consider whether or not  this source is a background AGN. 
Based on the log~$N$--log~$S$ of the \swift\ BAT AGN \citep{ajello12}, we expect 0.03 AGN at the flux limit of
6 $\times$ 10$^{-12}$~\ergcms\ in the 15--55 keV band
for the entire $D_{25}$ (3$^\circ \times 1^\circ$)
of M31.  If we restrict the area to the central 6\arcmin\ radius, the probability for an AGN is 5 $\times$ 10$^{-4}$, confirming
the estimate from \citet{revnivtsev14}.  
Similarly, we estimated the probability for the source being a background AGN  from the \nustar\ 8--24 keV band using the number counts from
\citet{harrison15} and obtained 0.05 and 8 $\times$ 10$^{-4}$ for the entire galaxy and the central 6\arcmin, respectively. 
These estimates suggest that  \bats\ is unlikely to be  a background AGN. 
In addition, the structure function  based on long-term variability  from 13 years of \cxo\ observations is not consistent with the ensemble  AGN structure function \citep{barnard14}, suggesting that \bats\ is unlikely to be a background AGN.
Furthermore, the shape of the X-ray spectrum having a flat slope would have to be a highly obscured AGN; 
however, we do not see a rising power law to 20--40 keV \citep[i.e.,][]{lansbury15,ptak15}.

The X-ray broadband (0.5--50 keV) spectral properties, i.e.,  the hard spectrum ($\Gamma \sim$ 1) at lower energies with a cutoff
around 15--20 keV, are more consistent with Galactic X-ray pulsars \citep[e.g.,][]{hung10,camero12,furst13}  than with black hole binaries and neutron star
 binaries with weak magnetic fields
(Z-sources and atoll sources). 
Our observations suggest that there is a soft X-ray excess in this system, and the soft excess has been also
seen in several  accreting pulsars \citep[e.g.][]{nagase02,hickox04}. 
One of the well-studied systems with a soft excess is the Galactic
X-ray pulsar  Her X-1.  In this case the soft excess is modeled as 
reprocessed hard X-rays through the inner edge of the accretion disk \citep{endo00,ramsay02}.  
The soft excess of Her X-1 is fit with a blackbody temperature of $kT_{\mathrm {bb}} = 0.09-0.12$ keV, which is similar 
to  the disk blackbody temperature found here for \bats.  The X-ray luminosity   of Her X-1
 is 3.1 $\times$ 10$^{37}$ \ergl\ \citep[in the 1.0 -- 50 keV band;][]{enoto07}.   
 If \bats\ belongs to M31, then the X-ray luminosity in the 1.0 -- 50 keV band is $\sim$
 4 $\times$10$^{38}$ \ergl, which is a factor of 10 higher than Her X-1.  However,
the bright end of the Galactic accreting pulsars reaches this luminosity (i.e., SMC X-1, LMC X-4, Cen X-3 and  RX J0059.2--7138).

Inspecting the UV and optical \hst\ images of the M31 bulge, we concluded that 
the potential optical counterparts of \bats\ have magnitudes that are inconsistent with the standard accretion disk model, and
so are likely unrelated to the accretion disk of \bats. 
Also,  the detected \hst\ matches are unlikely to be main sequence stars and probably less massive than 
3 \msun.   This suggests that the \bats\ system has either an evolved  1--3  \msun\  companion or a  $<$ 2 \msun\ main-sequence donor
below the \hst\ detection limit. 
The Her X-1 system also has an intermediate mass companion with 2 \msun\ (though a main sequence donor.  This makes it a unique system,
because a majority of accreting pulsars are known to be young systems having a Be, B, or O companion.  
\bats\ could be a very similar system to the Her X-1.  
\bats\ may not be a young system, but perhaps the compact object does not  have enough accreted material yet to decay its magnetic field,
making it still an X-ray pulsar.

Alternatively, \bats\ could be
a symbiotic X-ray binary.  This type of source has an M giant companion with
long spin and orbital periods, often found accreting via a wind. 
This type of source is
rare, with only 5 of them found in the Galaxy \citep{enoto14}.
 GX 1+4, the bright prototype, has an M giant donor with an upper limit of about 1.22 \msun\  \citep{hinkle06}, which
 may be comparable to the \hst\ matches we found for \bats.
The known  symbiotic X-ray binaries 
 have luminosity of 10$^{33}$ -- 10$^{36}$ \ergl, more than 2 orders of
magnitude lower than \bats.  However, it may be possible to brighten to $\sim$
10$^{38}$ \ergl\ depending on the wind velocity and the binary separation that would
determine the accretion rate.

Another counterpart of interest is a hot post-horizontal branch star candidate (the  blue source in the \hst\ image) in the region. 
However, it  is unlikely a
companion because it should be shrinking as it cools.  
We also note that \bats\ could be an ultracompact binary (i.e., 4U 1626-67 and 4U 1822-37)  with a white dwarf companion \citep{savonije86} that is under the \hst\ detection limit.
 This kind of system has a hard power spectrum below 10 keV \citep[see][]{esposito15}, which is similar to \bats. We point out that this type of
 system has a very short orbital period (on the order of subhours).  Detecting an orbital period and pulsation of \bats\ will certainly help determine the nature of the source.

Recently, \citet{esposito15} discovered the first accreting pulsar in the 
direction of an external arm of M31 with detection of its 1.2~s spin, thanks to its relatively high
pulse fraction ($\sim$50\%), using  \xmm\ observations.  
 They also found that its orbital modulation is about 1.3 days.  
The 0.3--10 keV luminosity is $\sim$10$^{37} $-- 10$^{38}$ ~\ergl, which is comparable to \bats.   Since there is no
potential high-mass donor in the field, they suggest that the system is likely an accreting pulsar with an intermediate
donor like Her X-1 or an ultracompact binary with a very low mass donor. This suggests that a similar population
to \bats\ exists in M31.  However, it is unclear why  only \bats\ dominates the entire galaxy at hard energies.

Lastly, we consider a possibility that it is a Galactic source along the line of sight toward M31, 
such as an Intermediate polar (IP) CV system, symbiotic X-ray
binary,  ultracompact binary or black widow. 
IP CV systems are known to have hard X-ray emission.  We can estimate a probability that  \bats\ is a Galactic IP using
the space density derived from the \swift\ BAT 70-month catalog by \citet{pretorius14}. 
Assuming that IPs are detectable up to 500 pc,
we expect 0.09 and 0.001 foreground IPs for the area of the entire M31 galaxy and the central 6\arcmin\ radius, respectively, for the observed 
 14--195 keV flux of 9 $\times$ 10$^{-12}$ \ergcms\ toward the M31 direction (b= $-$21$^{\circ}$).
The X-ray spectrum of these systems are generally characterized with thermal (multi-temperature) emission  with  Compton reflection with an association of the Fe K fluorescent line \citep[i.e.,][]{mukai15}.  
However,  the
\nustar\ observation does not show strong Fe emission lines.  We also applied a thermal plasma model instead of a power law
to the \nustar\ spectra, but the parameters are not well constrained ($F_{\rm 6.4 keV}$ $<$1.8$\times10^{-14}$ \ergcms\ for a 90\% upper limit  and  $kT >$ 80 keV which is exceeding the {\tt mekal} or {\tt apec} model limit), suggesting that it is unlikely an IP CV system.

If we assume a Galactic source located at 1--10 kpc,
the expected X-ray luminosity would be about 10$^{32}$ -- 10$^{34}$ \ergl\ in the 0.5--50 keV band.
At these distances,  the \hst\ image should identify its companion down to a
 0.2 -- 0.5 \msun\ main-sequence star.
The X-ray luminosity would be  
reasonable for  a Galactic symbiotic X-ray binary system \citep{enoto14},
but
the \hst\ image should identify its M giant companion. 
Hence, it is unlikely to be a Galactic symbiotic X-ray binary. 

Galactic black widows may contain very low mass companions ($<$ 0.1 \msun), which may be at or below the \hst\ 
 detection limit (e.g. $m_{\rm 814w} \sim$ 23 for PSR J1953+1846A at 4 kpc; \citealt{cadelano15}). Often, Galactic black widows are also known as  
 radio pulsars; however, there are no known radio sources with periodicities detected in the direction of M31 \citep{rubio13}. 
 The 0.3 -- 8.0 keV spectrum of 
Galactic black widows can be characterized by  blackbody plus power-law components  with similar photon index and $kT$ values \citep{gentile14} to
\bats. However, the flux of the thermal component is, in general, about 40\% (or higher) of the power law component  \citep{gentile14}.
In contrast, the thermal component of \bats\ is less than 15\% of the power law flux in the \swift\ data.  Also, the flux of \bats\ is about an order of magnitude higher.
   Therefore, the source is unlikely to be a Galactic black widow.

Some known Galactic ultracompact binaries (i.e., 4U 1626-67 and 4U 1822-37) also contain
 very low mass donors  ($<$ 0.1 \msun).  Unfortunately, it is difficult to compare from
 the Galactic ultracompact binary population, as it is not well sampled.
 Therefore, we cannot reject the possibility of \bats\ being a Galactic ultracompact binary completely.

Finally, the 0.5--2.0 keV spectrum of \bats\ suggests a disk-like feature, and often a disk is found in the bright end of the X-ray binary systems.  
We  conclude  that  \bats\ is likely to be an X-ray accreting pulsar with an intermediate-mass ($<$ 3~\msun) companion  or a symbiotic X-ray binary located  in M31 with X-ray luminosity of a few times 10$^{38}$~\ergl.
In either case, it dominates all emission from M31 at harder 
energies.  

\vspace{1cm}
We would like to thank the referee for his/her comments, which improved our manuscript. 
We sincerely thank  Neil Gehrels for approving the \swift\ DDT observations used
in this work.
We also thank \nustar\ and \swift\ mission planners for making the
 \swift\ and \nustar\ observations simultaneous.  
 This research has made use of the
   NuSTAR Data Analysis Software (NuSTARDAS) jointly developed by the ASI Science
   Data Center (ASDC, Italy) and the California Institute of Technology (Caltech, USA).
This work was supported by \nustar\ GO NNX15AV30G.
We are grateful to Antara Basu-Zych,  Hans Krimm, Craig Markwardt, Ryan Hickox,  Dheeraj Pasham,  Koji Mukai, Lennart van Haaften, and  Panayiotis Tzanavaris  for helpful discussions.  KLP acknowledges
funding from the UK Space Agency.
RB acknowledges funding from the Deutsches Zentrum f\"ur Luft- und
Raumfahrt grant 50 OR 1410.

Facilities: \nustar, \swift

\end{document}